# The metallization and superconductivity of dense hydrogen sulfide

Yinwei Li,[1*] Jian Hao,[1] Yanling Li[1], and Yanming Ma[2†]

[1]*School of Physics and Electronic Engineering, Jiangsu Normal University, Xuzhou 221116, P. R. China*

[2]*State Key Laboratory of Superhard Materials, Jilin University, Changchun 130012, P. R. China*

## Abstract

Hydrogen sulfide ($H_2S$) is a prototype molecular system and a sister molecule of water. The phase diagram of solid $H_2S$ at high pressures remains largely unexplored arising from the challenges in dealing with the looser S-H bond and larger atomic core difference between H and S. Metallization is yet achieved for water ice, but it was established for $H_2S$ above 96 GPa. However, the metallic structure of $H_2S$ remains elusive, greatly impeding the understanding of its metallicity and the potential superconductivity. We have performed an extensive structural study on solid $H_2S$ under high pressures through unbiased first-principles structure predictions based on swarm intelligence. Besides the findings of best-known candidate structures for nonmetallic phases IV and V, we are able to establish stable metallic structures violating an earlier proposal of elemental decomposition into sulfur and hydrogen [PRL **85**, 1254 (2000)]. Our study unraveled a superconductive potential of metallic $H_2S$ with an estimated maximal transition temperature of ~ 80 K at 160 GPa, higher than those predicted for most archetypal hydrogen-containing compounds (e.g., $SiH_4$ and $GeH_4$, etc).

Information on the structures of hydrogen-containing molecular systems at high pressures is central to many problems in physics, chemistry and allied sciences [1]. Due to the redistribution of electron density and the alterations of interatomic interactions of materials, a variety of fascinating physical phenomena have been observed or predicted for molecular compounds under pressure. Among these phenomena, the ubiquitous presence of a superconducting state in various high-pressure phases is attractive. In past decades, a large number of hydrogen-containing compounds (e.g., $SiH_4$[2-4], $SnH_4$[5, 6], $GeH_4$[7], $SiH_4(H_2)_2$[8], etc) have been predicted (or even observed [2]) to be superconductive at high pressures.

$H_2S$ is an analog of water at the molecular level. However, the phase diagram of solid $H_2S$ at high pressure is fundamentally different from that of water ice and is a challenge because of the looser S-H bonds and larger difference in atomic cores of S and H. At ambient pressure, $H_2S$ crystallizes in typically molecular solids (phases I-III, dependent upon temperature [9, 10]). Upon compression, $H_2S$ transforms into three high-pressure phases (phases IV, V, and VI) [11-14]). Due to the extremely weak X-ray scattering of hydrogen, the crystal structures of these high-pressure phases are under intensive debate [15-20]. Experimentally, tetragonal $I4_1/acd$ and monoclinic Pc structures have been suggested for phase IV by Fujihisa *et al.* [15] and Endo *et al.* [16], respectively. Theoretically, *ab* initio molecular dynamics (MD) simulations proposed three candidate structures for phase IV: partially rotational disordered tetragonal $P4_2/ncm$ [17], orthorhombic Pbca [18] and Ibca [19]. Pressurizing from an initial configuration of the $P4_2/ncm$ structure, MD simulations [20] predicted two orthorhombic structures of $Pmn2_1$ and $Cmc2_1$ as candidates for phases V and VI, respectively. The large structure diversity proposed for solid $H_2S$ posts great challenges in understanding of its high-pressure phase diagram. Further investigation on the high-pressure structures of $H_2S$ is greatly demanded.

The most fascinating topic about $H_2S$ is the metallization at moderate pressures (~96 GPa). This is in contrast to the situation in water ice where decomposition into $H_2O_2$ and a hydrogen-rich $H_{2+\delta}O$ ($\delta \geq 1/8$) compound at terapascal pressures regions [21] was predicted before the metallization. Optical experiments [13] showed that the color of $H_2S$ changes from thin yellow to black at ~ 27 GPa, indicating a large decrease of the energy gap. An infrared spectral study [14] demonstrated that $H_2S$ eventually turns into a metal at 96 GPa. It is essential to uncover the metallic structure in order to understand the metallicity of $H_2S$ and its potential superconductivity. To note, the metallization has been interpreted as the metallization of elemental sulfur rather than the compound, since $H_2S$ was predicted to decompose into sulfur and hydrogen under the metallic pressure [20]. We show in this work by first-principles structure predictions that $H_2S$ is stable against this decomposition at least up to 200 GPa and the metallization is in close correlation with the actual bandgap closure.

Structure searching simulations for $H_2S$ were performed through the swarm intelligence CALYPSO structure prediction method [22, 23] as implemented in the CALYPSO code [22]. This approach has been successfully applied into investigation of a large variety of compounds at high pressures (e.g., Refs. [24-29]). The underlying *ab initio* structure relaxations were performed using density functional theory within the Perdew-Burke-Ernzerhof generalized gradient approximation as implemented in the Vienna *ab initio* simulation package (VASP)[30]. The all-electron projector augmented wave (PAW)[31] method was adopted with the PAW potentials taken from the VASP library. For the structure searches, an energy cutoff of 700 eV and Monkhorst-Pack Brillouin zone sampling grid with the resolution of $2\pi \times 0.05$ Å$^{-1}$ was used. The obtained structures were re-optimized more accurately with grids denser than $2\pi \times 0.03$ Å$^{-1}$ and energy cutoff of 1000 eV, resulting in total energy convergence better than 1 meV/atom. Phonon dispersion and electron-phonon coupling (EPC) calculations were performed with density functional perturbation theory using the Quantum-ESPRESSO package [32] with a kinetic energy cutoff of 90 Ry. $4\times4\times3$ and $6\times6\times2$ *q*-meshes in the first Brillouin zones were used in the EPC calculations for the P-1 and Cmca structures, respectively.

We performed structure predictions with simulations cells containing 1–8 formula units (f.u.) under pressures in the range 10 - 200 GPa. Interestingly, within the whole pressure range studied, we did not find any previously proposed structures, but instead five completely new low-enthalpy structures with space groups of $P2_12_12_1$, $Pc$, $Pmc2_1$, $P-1$ and $Cmca$ (Fig. 1 and Ref. [33] for detailed structure information) were uncovered. Phonon dispersion calculations of these structures do not give any imaginary frequencies and therefore have verified their dynamical stabilities [33]. The enthalpy curves of these new structures together with the ambient phase III [10] and the previously proposed structures were plotted as a function of pressure in Fig. 2 (a). According to our results, the known high pressure phase diagram should be fundamentally revised since our predicted structures are energetically more stable than all earlier structures at the entire pressure ranges studied. We found that ambient-pressure phase III transforms to the $P2_12_12_1$ structure at 12 GPa and then to the $Pc$ structure at 28 GPa. Both $P2_12_12_1$ and $Pc$ structures consist of four $H_2S$ molecules per unit cell, forming typical van der Waals molecular solids (Fig. 1 a, b). The averaged intra-molecular S-H bond lengths are 1.367 Å at 20 GPa and 1.385 Å at 30 GPa for the $P2_12_12_1$ and $Pc$ structures, respectively, slightly longer than that (1.336 Å) in a gas $H_2S$ molecule.

With increasing pressure, the polymerization of $H_2S$ takes place with the evolution of the $Pc$ structure. As shown in Fig. 1 (b), half of hydrogen atoms move toward to the midpoints of two neighboring S atoms. Consequently, one-dimensional chains of edge-sharing $SH_3$ tetrahedrons (each S forms three covalent bonds with three H atoms) are formed and the $Pc$ structure eventually transforms to the $Pmc2_1$ structure at ~ 65 GPa without the major modification of S sublattice (Fig. 1c). The polymerization behavior of $H_2S$ is rather unique and is in clear contrast to the hydrogen bond symmetrization in $H_2O$ taking place above 60 GPa [14, 34], where each oxygen atom forms four covalent bonds with four H atoms. The polymerization mechanism predicted here differs from the earlier proposal [14], in which H atoms were squeezed out of the hydrogen-bond axes into interstitial spaces forming various S-S bonds. The available experimental X-ray diffraction data for phases IV and V of $H_2S$ are generally low in quality, consisting of several main broad peaks [33]. Our predicted $P2_12_12_1$ and $Pc$ ($Pmc2_1$) structures are able to reproduce these main diffraction peaks for phases IV and V, respectively [33]. However, for conclusive structural solutions of phases IV and V, finer experimental data are required.

Upon further compression, the $Pmc2_1$ structure transforms to P-1 structure at 80 GPa (Fig. 2a). The P-1 structure is intriguing with various S-S bonds, forming dumbbell-like $H_3S$-$SH_3$ units. It is worth noting that the $H_3S$-$SH_3$ units are linked to each other through planar symmetric hydrogen bonds via H2 and H3 atoms (Fig. 1), forming edge-sharing planar $S_6H_4$ quasi-rectangles (Fig. 1d and f). Two inner H1 atoms per $S_6H_4$ quasi-rectangle do not form bonds within the layer, but bond with S atoms in adjacent layer. Therefore, the P-1 structure is a typical three-dimensional structure.

The P-1 structure is stable up to 160 GPa, above which the Cmca structure (Fig. 1e) takes over the phase stability. The Cmca structure is rather similar to the P-1 structure in a manner that Cmca structure is also composed of $S_6H_4$ quasi-rectangles and its interlayers are bridged through S-H1 bonds. However, different from the one directional arrangement in the P-1 structure, the $S_6H_4$ quasi-rectangles in the Cmca structure distribute in two orientations, as shown in Figs. 1 (f) and (g). Interestingly, each S atom of Cmca structure bonds with two H1 atoms in adjacent two layers, resulting in 5-fold coordination of S atom. More intriguingly, all H atoms in the Cmca structure become bonded with two S atoms. Therefore, the Cmca phase of $H_2S$ can be recognized as an "atomic crystal".

The decomposition enthalpies into elemental sulfur and $H_2$ have been calculated to examine the phase stability of $H_2S$ under pressure (Fig. 2a). By assuming the earlier $Cmc2_1$ structure [20], our calculations indeed revealed the elemental decomposition beyond 80 GPa, in excellent agreement with the previous results. However, with the findings of our new structures, the $Cmc2_1$ structure is never stable and the proposed decomposition scenario [20] should be ruled out. It is worth noting that the zero-point energy (ZPE) may play an important role in determining the phase stability of hydrogen-containing compounds due to the small atomic mass of H atom. Our calculations showed that the inclusion of ZPE does not change the phase sequence as depicted in Fig. 2 (a). For example, the ZPE effects do not modify the decomposition enthalpy since the ZPE value (6 meV/f.u. at 160 GPa) for the decomposition is rather small, hardly comparable to the enthalpy data (249 meV/f.u.) at the same pressure as derived from calculations by assuming a static lattice.

Our band structure calculations [33] show that $H_2S$ is an insulator at ambient pressure with a large band gap of ~ 4.2 eV, in agreement with experimental observations. With increasing pressure, the band gap decreases gradually to ~ 3 eV at 15 GPa in $P2_12_12_1$ phase and is further reduced to ~ 1.3 eV at 30 GPa in Pc phase. Such a small gap in the band structure of the Pc phase could naturally explain the color change at ~ 27 GPa observed experimentally, where $H_2S$ becomes black but not metallic [13]. The band gap closure occurs with the P-1 phase at ~ 110 GPa [33], a pressure slightly higher than the experimental metallization pressure (96 GPa) [14]. $H_2S$ becomes a good metal around 120 GPa having a large electron density of states at Fermi level ($N_F$, 0.3 eV$^{-1}$ per f.u.) (Fig. 2b). The $N_F$ value has a dramatic increase at the transition to the atomic Cmca phase (0.51 eV$^{-1}$ per f.u. at 160 GPa) and then decreases slightly with pressure (the inset in Fig. 4a). Band structures for both the P-1 and the Cmca phases feature "Flat-Deep" band characters, which may suggest their superconductive potentials [34].

Figure 3 presents the phonon band dispersions and partial phonon density of states of the P-1 and Cmca phases at selected pressures. As expected, the low frequencies (< 15 THz) are dominated by the vibrations of S, whereas the high end of the spectrum is due to the H atoms. Under pressure, we find interesting features in our computed phonon dispersion curves. For the P-1 structure, most phonons harden with pressure; however, the phonon modes around Γ and Z points at ~ 20 THz and along the F-Q direction at ~ 50 THz soften with pressure (Fig. 3a). Subsequent analysis showed that these phonon softenings play a crucial role in enhancing the electron-phonon coupling of P-1 phase under pressure. For the Cmca structure, all phonon modes harden up with pressure.

The Eliashberg spectral function $\alpha^2F(\omega)/\omega$, logarithmic average frequency $\omega_{\log}$ and electron-phonon coupling parameter λ for the P-1 and *Cmca* structures are explicitly calculated to explore the potential superconductivity of $H_2S$ (Fig. 4). The superconducting $T_c$ was estimated by using the Allen and Dynes modified McMillan equation [35] with a typical choice of $\mu^* = 0.13$. It is found that once $H_2S$ is metalized, it has a high potential to be superconductive. The calculated $T_c$ increases linearly with pressure in the P-1 phase from 33 K at 130 GPa to 60 K at 158 GPa (main panel of Fig. 4a). Intriguingly, $T_c$ increases abruptly up to 82 K at the transition to the Cmca phase, and then decreases monotonically to 68 K at 180 GPa. From Fig. 4 (b), we found that the low frequency S vibrations in P-1 phase contribute to 47% of the total λ, while the remaining 53% is from the H vibrations. As pressure increases to 150 GPa, three distinct peaks of spectral functions at 15, 20 and 46 THz appear in P-1 phase (Fig. 4b), arising from pressure-induced phonon softening. These phonon softenings lead to a pronounced increase of λ from 0.77 at 130 GPa to 0.97 at 150 GPa. The substantial increase of $T_c$ in the Cmca phase mainly arises from the sharply elevated $N_F$, while the phonon hardening gives rise to the decreased λ and thus a lowering of $T_c$.

In a survey of the literature, we find that superconductivity is generally predicted in hydrogen-containing compounds with stoichiometry AH or $AH_x$ ($x \geq 3$). $H_2S$ is peculiar since it is the first $AH_2$-type example having the potential to be superconductive. The estimated maximal $T_c$ of 82 K in $H_2S$ is rather high, even higher than those predicted for most archetypal hydrogen-containing compounds, such as $SiH_4$, $GeH_4$, $YH_3$, and $ScH_3$, etc.

Our current work has demonstrated that the precise determination of structures is crucially important to understand the physical properties of materials. The use of the advanced structure prediction methods (here CALYPSO method) unbiased by any prior known structural knowledge allows us to locate better structures than those previously proposed by other simulations [17-20]. Indeed, as we show here, earlier decomposition picture of $H_2S$ was fundamentally revised.

In summary, we have performed extensively first-principles structure searching calculations on $H_2S$ in a large pressure regime up to 200 GPa. Five high-pressure phases were predicted and are found to be energetically more stable than all earlier structures at certain pressure ranges, leading to a fundamental modification of the entire high-pressure phase diagram of $H_2S$. Our predictions have led us to conclude that $H_2S$ is thermodynamically stable with respect to elemental decomposition into sulfur and hydrogen at least up to 200 GPa. This result stands in sharp contrast with earlier proposal on compositional instability of $H_2S$ and revises our understanding on the observed metallicity of $H_2S$. Moreover, our electron-phonon coupling calculations established the superconductive potential of the metallic $H_2S$. Our work has provided a key step forward for understanding the high-pressure phase diagram of solid $H_2S$ and will inevitably stimulate future experiments to explore the superconductivity on a sister compound of water ice, whose superconductivity becomes impossible.

Y. L., J. H., and Y. L. acknowledge the funding supports from the National Natural Science Foundation of China under Grant Nos. 11204111 and 11047013, the Natural Science Foundation of Jiangsu province under Grant No. BK20130223, and the PAPD of Jiangsu Higher Education Institutions. Y. M. acknowledges the funding supports from China 973 Program under Grant No. 2011CB808200, and the National Natural Science Foundation of China under Grant Nos. 11274136, 11025418, and 91022029.

*yinwei_li@jsnu.edu.cn

†mym@jlu.edu.cn

# Figure captions

**Fig. 1.** (color online) The energetically favorable structures predicted by CALYPSO structure searching for $H_2S$: (a) molecular $P2_12_12_1$ structure, (b) molecular Pc structure, (c) polymeric $Pmc2_1$ structure, (d) polymeric P-1 structure and (e) atomic Cmca structure. Arrows in (b) illustrate the displacements of H atoms under pressure. The inset figures in (d) and (e) represent the bonding situation of each S atom. (f) and (g) show the charge densities within layer in the P-1 (90 GPa) and Cmca (160 GPa) structures, respectively.

**Fig. 2.** (color online) (a) Enthalpy curves of various structures with respect to the previously predicted Ibca structure [19]. Note that the results of some previously proposed structures are not shown since their enthalpies are extremely high, out of comparison range in the figure. The decomposition enthalpy into $S+H_2$ was also plotted, where the structures of phases I [36], III [37], IV [38], and V [38] for S and $P6_3/m$ and $C_2/c$ structures for $H_2$ [39] were adopted, respectively. Inset in (a) is an enlarged view of the phase transition sequence in pressure range 10 - 70 GPa. (b) and (c) are the electronic band structures and total density of states (DOS in unit of $eV^{-1}$ per f.u.) for P-1 structure at 120 GPa and Cmca structure at 160 GPa, respectively.

**Fig. 3.** (color online) (a) and (b) represent phonon dispersions, partial phonon density of states (PHDOS) for P-1 at 130 GPa and Cmca at 160 GPa, respectively. Red dashed lines in left panel of (a) show the softening phonon modes in the P-1 phase when pressure is increased up to 150 GPa.

**Fig. 4.** (color online) (a) Pressure dependence of $T_c$ within P-1 and Cmca phases. Insets in (a) show the evolutions of the $N_F$ (left panel) and the logarithmic average phonon frequency ($\omega_{log}$) and λ (right panel) with pressure. (b) and (c) show the spectral functions $\alpha^2F(\omega)/\omega$ and electron-phonon coupling integration of $\lambda(\omega)$ at selected pressures for P-1 and Cmca structures, respectively. Shaded regions in (b) show the significant contribution of three strong peaks of $\alpha^2F(\omega)/\omega$ to λ.

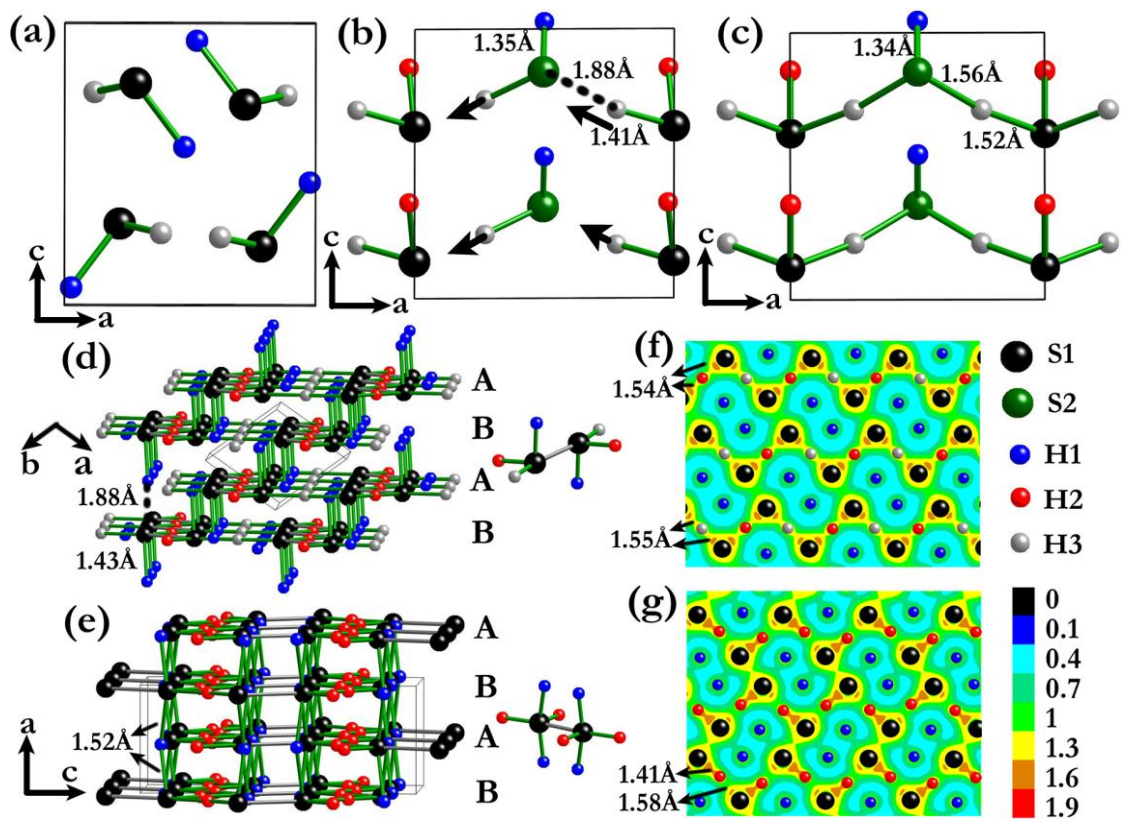

Fig. 1

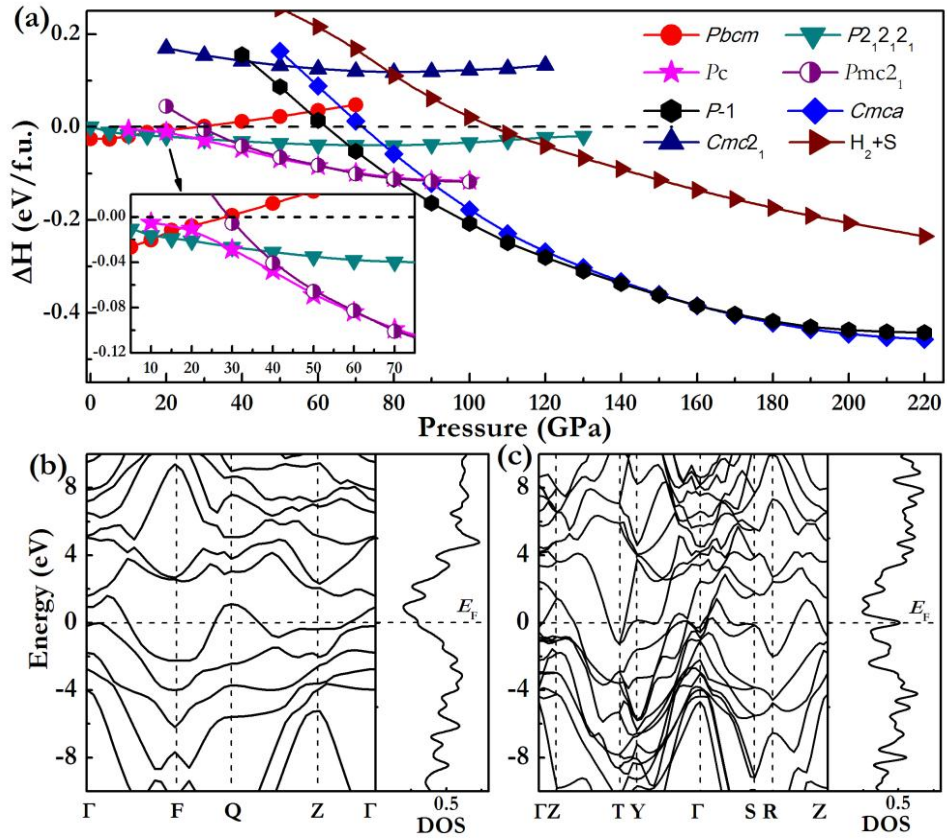

**Fig. 2**

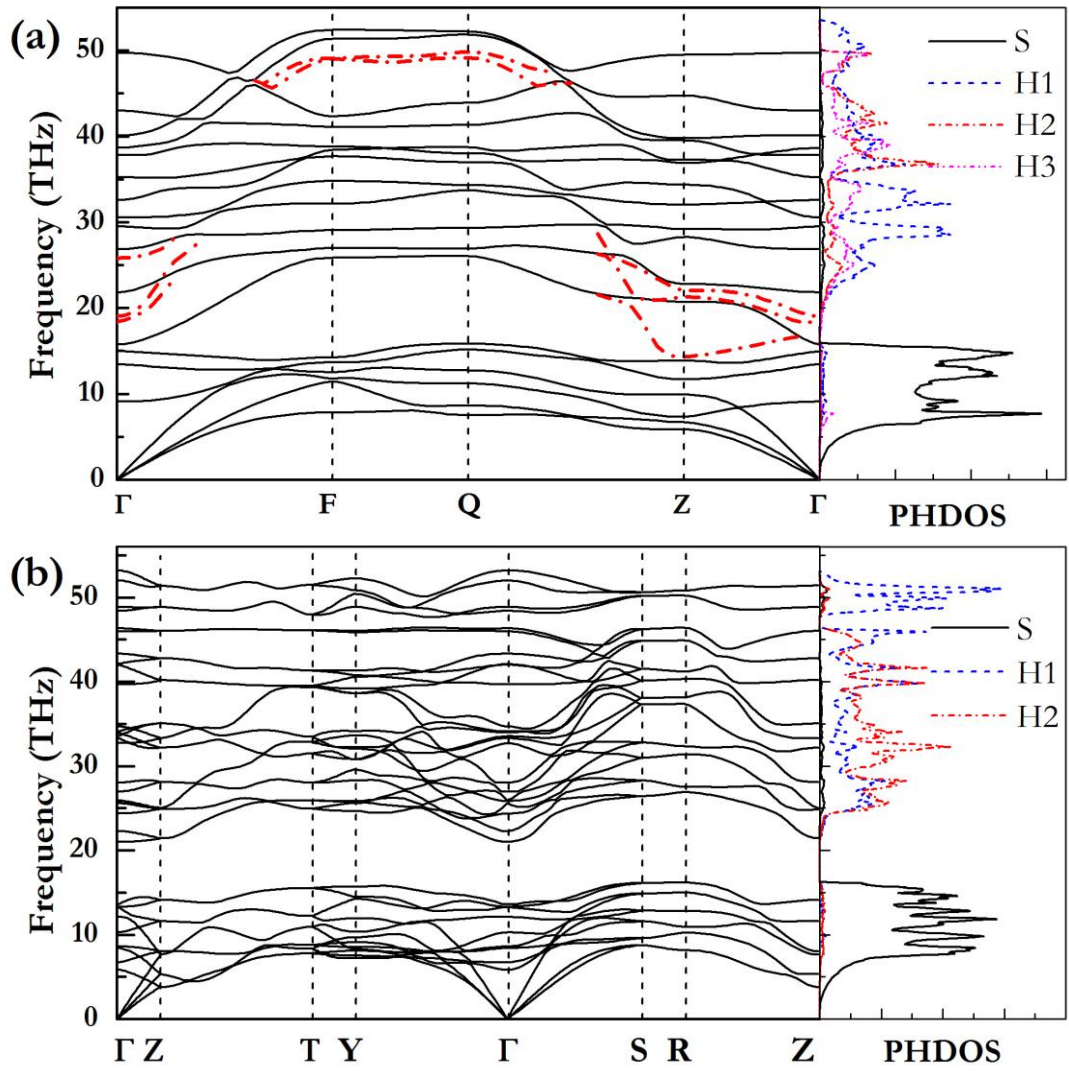

Fig. 3

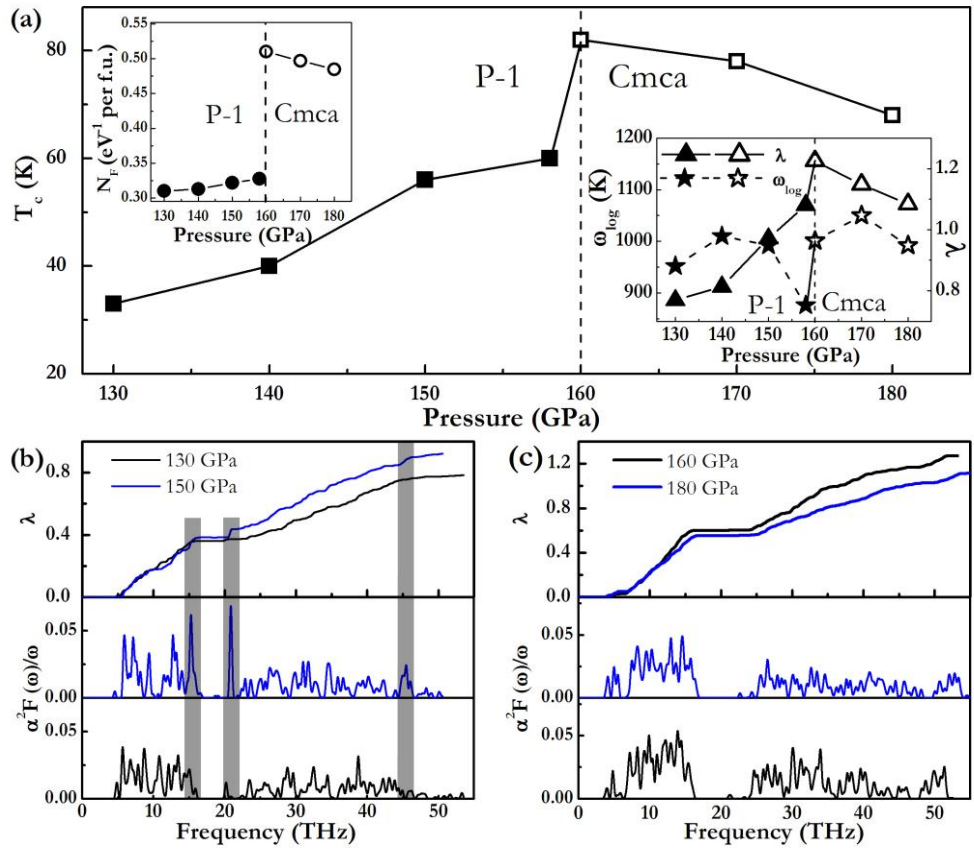

**Fig. 4**